\documentclass[aps,pre,twocolumn,showpacs,showkeys,floatfix]{revtex4}
\usepackage{graphicx}
\usepackage{longtable}
\usepackage{dcolumn}
\usepackage{bm}
\begin{document}
\preprint{IFT/14/06}
\title{ Limit cycles of effective theories }
\author{ Stanis{\l}aw D. G{\l}azek }
\affiliation{ Institute of Theoretical Physics, Warsaw University, 
              ul.  Ho{\.z}a 69, 00-681 Warsaw, Poland } 
\begin{abstract}
A simple example is used to show that renormalization group limit 
cycles of effective quantum theories can be studied in a new way. 
The method is based on the similarity renormalization group procedure 
for Hamiltonians. The example contains a logarithmic ultraviolet 
divergence that is generated by both real and imaginary parts of the 
Hamiltonian matrix elements. Discussion of the example includes a 
connection between asymptotic freedom with one scale of bound states 
and the limit cycle with an entire hierarchy of bound states.
\end{abstract}
\label{sec:intro}
\maketitle
\section{ Introduction }

A model~\cite{model} invented in the front form of
Hamiltonian dynamics~\cite{Dirac1,Dirac2} has been used in
its generic matrix version to
argue~\cite{GlazekWilsonExample} that cyclic dependence of
coupling constants on cutoffs in renormalization group (RG)
procedure~\cite{WilsonLimitCycle} may actually be commonplace
for quantum mechanical systems whose Hamiltonians require
renormalization. In independent studies, many examples of
physical few-body systems with short range interactions and
large scattering length have been considered in which a
cycle structure may occur with universal
features~\cite{BraatenHammerReview}. In the theory of
three-body systems in nuclear~\cite{BedaqueHammerKolck} and
atomic~\cite{BraatenHammer} physics, the limit cycle is
associated with the existence of a sequence of bound states
with binding energies forming a geometric series. The cycle
has also been discussed in the context of the Bose-Einstein
condensation~\cite{BraatenHammerKusunoki} and
superconductivity~\cite{LeClairRomanSierra}. Even the
elementary non-relativistic Hamiltonian for one particle in
the potential $\sim 1/r^2$, when properly regulated,
exhibits the cyclic behavior~\cite{BraatenPhillips}.
Concerning particle theory, the cyclic behavior of effective
nuclear interactions could occur if the masses of
$\pi$-mesons, which influence the shape of nuclear
potentials, were increased by about one third. Since the
masses of $\pi$-mesons can be linearly related to the masses
of up and down quarks using the Gell-Mann-Oakes-Renner
relation, the nuclear theory suggests that the masses of
light quarks in the Standard Model are near the range where
QCD may develop an infrared (IR) limit
cycle~\cite{BraatenHammerQCD,EpelbaumHammerMeissnerNogga}.
In order to determine if the IR limit cycle does indeed 
occur in QCD, the binding mechanism for quarks and
gluons will have to be understood much more precisely than 
it is so far. While the mechanism of binding in just
three-body systems has fascinated researchers for many
years~\cite{Thomas,Efimov} and a large body of literature
exists on the subject~\cite{NielsenFedorovJensenGarrido},
studies of the RG limit cycle appear to be in their
infancy~\cite{GlazekReview}. Quantum three-body calculations
that trace the dependence of coupling constants on cutoffs
tend to be complex~\cite{MohrFurnstahlPerryWilson}. Thus,
the simple matrix example~\cite{GlazekWilsonExample}
provides useful insights concerning universality in
Hamiltonian quantum mechanics with RG limit
cycle~\cite{GlazekWilsonUniversality}. 

So far, all studies of limit cycles known to the author have
been carried out in one of two ways. The first way,
reminiscent of the concept of charge renormalization in
quantum electrodynamics~\cite{GellMannLow}, is to regulate a
diverging model with a cutoff and solve for some observable,
such as a scattering amplitude or a bound-state energy,
which is postulated not to depend on the cutoff. When the
cutoff parameter is varied, it is possible to keep the
observable fixed if instead a coupling constant is allowed 
to vary (there may be more than one constant to vary). Limit cycle
means that the coupling constant is a periodic function of
the cutoff parameter when the latter increases to infinity.
Most of the quoted literature discusses examples of a limit
cycle in this way. But if it is known that a theory being
considered is only applicable to phenomena of a limited
range of scales, the limit of infinite cutoff becomes
logically questionable. The limit may not exist if the
required coupling constants diverge in or on the way to the
limit. The question of how to guarantee that all observables
of interest are independent of a changing cutoff when it is
finite (instead of being infinite) is hard to answer
proceeding along the first way (unless one deals with a
model in which the introduction of a small number of finite
cutoff-dependent coupling constants is sufficient to
completely eliminate the cutoff from equations that describe
the model).

The second way is to apply Wilson's renormalization group
procedure to a diverging model with a cutoff and compute the
number of required constants and their shapes as functions
of the cutoff~\cite{WilsonT1,WilsonT2}. The required number
of constants may be very large, or infinite, which is a
typical situation for finite cutoffs (see below). The
calculation involves a reduction of the number of degrees of
freedom that are explicitly kept in the dynamics, and an
evaluation of the change of the dynamics due to the
reduction. Precise calculations can be done using discrete
degrees of freedom and the number of the relevant degrees of
freedom can be kept finite. The boundary of the set of the
retained degrees of freedom plays the role of a new finite
cutoff that is much smaller than the cutoff of the initial
regularization (in the case of fixed-point criticality,
finite cutoffs only mean that the system can be described by
the same model at all scales). The finite cutoff defines the
domain of an effective theory. If the calculation is
precise, then by construction the finite boundary cannot
have any effect on the solution and can be set arbitrarily.
However, the finite boundary limits the effective theory to
a small space where the only source of dependence on the
initial regularization cutoff is the explicit, computable
dependence of the matrix elements of the effective-theory
Hamiltonian on the initial cutoff. Using this information,
it is possible to identify the changes in the setup of the
initial regulated model that guarantee that the Hamiltonian
matrix elements in effective theories with finite cutoffs do
not depend on the initial cutoff. All predictions of such
effective theory are then independent of the initial cutoff.
It may happen that the presence of new terms, called
counterterms, amounts to a replacement of the initial
coupling constants with the new ones that depend on the
initial cutoff. In principle, the outcome of the procedure
is a family of effective theories with finite cutoffs and
predictions that are guaranteed to be independent of the
finite cutoffs. If a complete theory must contain many
constants depending on the finite cutoff, additional
conditions, such as the requirement of
symmetries~\cite{PerryReview}, may lead to finite
correlations among otherwise unconstrained finite constants
in the effective theories. One can consider the limit of an
infinite number of degrees of freedom by simultaneously
increasing the initial cutoff and using the RG
transformation over an increasingly large range of scales to
obtain effective theories with the same size of finite
cutoffs. The limit cycle is then the ultimate RG orbit on
which the effective coupling constants move indefinitely in
the case of the infinite number of initial degrees of
freedom. To approximate the limit cycle with increasing
precision when the number of full cycles in the RG procedure
increases, the initial regulated model must be set up in
such a way that all unwanted terms that cause departures
from or impede approach to the limit cycle in the RG
procedure are eliminated~\cite{GlazekWilsonUniversality}.

This article concerns a new way to describe the cyclic
behavior of effective theories, closely related to but
different from the two ways mentioned above. Namely, the
Hamiltonian matrices of effective theories are calculated
using a renormalization group procedure called
similarity~\cite{GlazekWilsonSimilarity1,
GlazekWilsonSimilarity2}. In the new procedure, no states
are eliminated from the effective dynamics. Instead, the
basis states are changed in such a way that the effective
interaction Hamiltonian can dynamically couple only those
basis states whose kinetic energies differ by less than an
arbitrarily prescribed scale $\lambda$. This scale plays the
role of a new renormalization group parameter. $\lambda$
varies from infinity in the initial regulated model down to
a finite value in an effective theory. When the initial
regularization is being relaxed, the Hamiltonian matrix
evaluated in the effective basis corresponding to a finite
$\lambda$ can lead to a dependence of perturbatively
calculated observables on the initial regularization only
through the explicit dependence of the Hamiltonian matrix
elements on the initial cutoff. This explicit dependence is
used to find the counterterms in the Hamiltonian of the
initial model. Once the counterterms are found, the
effective theories are parametrized by finite values of
$\lambda$ and all have the same predictions that are
independent of the initial regularization. The dependence of
the effective Hamiltonian matrices obtained using the
similarity RG procedure on $\lambda$ is described in the 
following sections in the case of the same generic example 
that was previously studied in Refs.~\cite{GlazekWilsonExample}
and~\cite{GlazekWilsonUniversality} using other methods.

The similarity procedure requires a generator (see next
section) and details of the cyclic behavior of effective
theories depend on how the generator is chosen. Since the
main features of the cycle that are related to the spectrum
of a renormalized model are not sensitive to the choice of
the generator, variations in the cyclic behavior due to
variations in the similarity generator will not be discussed
in any detail. The explicit discussion in this article is
focused on the case of a very elegant generator taken from
Wegner's flow equation for (partial, in the case of a
degenerate spectrum) diagonalization of Hamiltonian
matrices~\cite{Wegner1,Wegner2}. Similarity RG procedures
using other generators may have better convergence
properties in perturbative
calculations~\cite{GlazekMlynikAlteredWegner}, but this is
of no significance to a qualitative understanding of the
limit cycle whose description here is based on a
non-perturbative numerical solution. The model considered
here does not have degenerate eigenvalues and Wegner's
generator leads to complete diagonalization when the RG
procedure is carried out down to energy scales below the
smallest eigenvalue.

Wegner~\cite{Wegner3} has recently pointed out that
differential equations similar to his had been considered
earlier for various purposes (other than a RG procedure) by
mathematicians~\cite{ChuDriessel, Brockett,Chu}. The study
described here is limited to the context of RG
procedure~\cite{GlazekWilsonSimilarity1,
GlazekWilsonSimilarity2}. The study shows that the limit
cycle behavior of effective theories resembles in part the
behavior of asymptotically free ones. Since the similarity
approach can be applied to gauge theories (e.g., see
\cite{LFQCD} and later literature on the light-front
formulation of QCD), the resemblance observed here in the
simple model is also of interest in particle physics.

The model studied here is so simple that the cyclic
dependence on $\lambda$ of a large number of matrix elements
of effective Hamiltonians can be approximately described in
terms of just one function of $\lambda$. This function is
called the renormalized coupling constant, $g_\lambda$,
because it corresponds to and is closely related to
renormalized coupling constants in effective Hamiltonians of
quantum field theories. The main result is that each and
every RG cycle consists of a sequence of gradual changes of
$g_\lambda$, like in asymptotically free theories, followed
by a sudden large change that brings the coupling constant
back to its initial value in the cycle. In principle, this
pattern repeats itself indefinitely when $\lambda$ changes
indefinitely. Each new cycle corresponds to a new bound
state whose binding energy is on the order of $\lambda$ at
which the sudden change occurs. 

The paper is organized as follows. Section \ref{sec:model}
describes the model and produces the cyclic behavior of 
$g_\lambda$. Section \ref{sec:change} explains some elements 
of the mechanism of change that $g_\lambda$ exhibits in 
every cycle. Section \ref{sec:discussion} concludes the 
paper with comments concerning implications of the model 
study.
\section{ Model }
\label{sec:model}
The model Hamiltonian is constructed for the purpose of 
qualitative understanding of the logarithmic similarity RG 
effects in the presence of bound states, which is a typical 
situation in the case of RG limit cycles. In order to 
construct a model that can be solved precisely, each 
and every energy scale is represented by just one state.
The energy scale is defined by the spectrum of a Hamiltonian
$H_0$ whose eigenvalues are assumed to be given by the formula 
$E_n = b^n$, where $b>1$ and $n$ is an integer. This means that 
the spectrum of $H_0$ is equally spaced on the logarithmic 
scale with step $\ln {b}$ (the unit of energy is set to
1). The corresponding eigenstates, denoted by $|n\rangle$, 
form an orthonormal basis, $\langle m| n\rangle = \delta_{mn}$. 
The interaction Hamiltonian in the model, $H_I$, is defined 
by its matrix elements in this basis. 

Building models with such discrete basis is not a new 
idea \cite{WilsonT1,WilsonT2}. It is also a starting point for 
consideration of the case of $b \gg 1$. The case of large $b$ 
is not accurate physically but it can further simplify the 
mathematics. There may exist complex physical systems where 
one can gain new insights by considering the case of very 
large $b$. For example, such strategy was instructive in the 
case of the Kondo problem \cite{WilsonKondo}, although 
quantitative numerical results were obtained for $b$ around 2. 
There is hope that some way can be found to make use of a large 
$b$ in studies of light front theory \cite{LFQCD}. There, one 
also needs to find ways to drastically reduce the number of 
discrete degrees of freedom in dimensions other than the energy 
dimension. 

The initial model Hamiltonian, $H=H_0 + H_I$, is defined 
by its matrix elements \cite{GlazekWilsonExample}
\begin{equation}
H_{mn}(g,h) = (E_m E_n)^{1/2}
\left[\,\delta_{mn} - g - i h s_{mn}\right] \, .
\label{h}
\end{equation}
For $m=n$, $\delta_{mn}= 1$ and $s_{mn} = 0$. For $m \neq n$, 
$\delta_{mn}=0$ and $s_{mn} = (m-n)/|m-n|$. The interaction 
Hamiltonian, $H_I = H - H_0$, vanishes when $g=h=0$. 

The Hamiltonian of Eq.
(\ref{h}) has a generic ultraviolet structure with a
logarithmic divergence to which both real and imaginary
parts of the matrix elements contribute. Initial
regularization is imposed by the cutoff $E_n \le \Delta$.
(In \cite{GlazekWilsonExample}, the same cutoff was denoted
by $\Lambda$. The notation is changed to avoid confusion
with $\lambda$ to be introduced below.) If one sets $\Delta
= b^N$, the cutoff means $n \le N$ and a logarithmic
divergence in $\Delta$ means a linear divergence in $N$.
When $h=0$, the perturbative algebraic similarity
procedure~\cite{GlazekWilsonSimilarity1} produces
interaction Hamiltonian with matrix elements of the form
$g_\Delta \, (E_m E_n)^{1/2}$. This means that for $h=0$ the
counterterm that removes the dependence on $\Delta
\rightarrow \infty$ from the model eigenvalues is obtained
by replacing the constant $g$ with a cutoff-dependent
coefficient $g_\Delta$. In fact, $g_\Delta$ exhibits
asymptotic freedom as a function of $\Delta$ when $h = 0$.
When $h \neq 0$, $g_\Delta$ exhibits a limit cycle behavior
(or chaos)~\cite{GlazekWilsonExample}. $h$ does not depend
on $\Delta$.
 
The similarity procedure yields a family of effective 
theories with Hamiltonian matrices labeled by $\lambda$. 
The family is described by a solution to the equation 
of the form
\begin{equation}
\label{srg}
 {d \over d \lambda} \, H_\lambda \, = \, 
 \left[ F_\lambda \{ H_\lambda \} , H_\lambda \right] \, ,
\end{equation}
where $F_\lambda \{ H_\lambda \}$ denotes the generator of
the similarity transformation. The initial condition is set
at $\lambda = \infty$ (in a numerical study, one can work
with inverse of $\lambda$ and start from 0, or the infinity
is replaced by any number much greater than $\Delta$). In
the model studied here, $ H_\infty$ is given by $H$ in Eq. 
(\ref{h}) with $g$ replaced by $g_\Delta$. The case with 
$h=0$ has been originally studied using Wegner's form of 
the generator $F_\lambda \{ H_\lambda \}$ in 
\cite{WilsonGlazekAustralia,GlazekWilsonAFBS}. Writing 
the generator $F_\lambda \{H_\lambda\}$ in terms of its 
matrix elements as
\begin{equation}
\label{generator}
\langle m| F_\lambda\{H_\lambda\} |n\rangle \, = \, 
f_{mn} (D_m - D_n) H_{\lambda m n} \,\, ,
\end{equation}
where $D_m = H_{\lambda m m}$ is the diagonal matrix
element number $m$ of $H_\lambda$, one obtains the following 
set of equations for all matrix elements of $H_\lambda$,
\begin{eqnarray}
\label{dhmn}
& & {d \over d \lambda} \, H_{\lambda m n}  = \,  \\ 
& & \sum_{k = M}^N  [f_{mk}(D_m - D_k) + f_{nk}(D_n - D_k)] 
                H_{\lambda m k} H_{\lambda k n} \, .
\nonumber 
\end{eqnarray}
The lower bound on the summation range, $M$, is an additional 
parameter introduced here to simplify numerical analysis of the 
low energy region. It sets an infrared bound on the lowest 
energy included in the model, $E_M = b^M$. The lower bound is 
very small for large negative values of $M$. The actual size 
of $M$ is irrelevant to almost all aspects of the limit cycle 
behavior discussed here. The only exception are practical numerical 
issues and the range of $\lambda$ over which one can easily 
observe the cycle using personal computers. $M$ was set to $-25$ in 
all numerical examples described below. 

Wegner's flow equation~\cite{Wegner1} is obtained for $f_{mn} = 
d \mathit{l}/d\lambda = -2/\lambda^3$. This choice amounts 
to the change of the RG variable $\lambda$ to Wegner's
parameter $ \mathit{l} = 1/\lambda^2$.

The behavior of $(N-M+1)^2$ matrix elements of $H_\lambda$
as functions of $\lambda$ results from solving the coupled
nonlinear differential Eqs. (\ref{dhmn}). Numerical
experiments showed that in order to clearly discern
logarithmic effects for small coupling constants $g_\Delta$
and $h$ (on the order of 0.1 or smaller) using a contemporary
personal computer for less than an hour, one needs about 30
basis states with $b$ on the order of $2$. This gives about
1000 functions to trace. This number is sufficient for
finding dominant features in the similarity RG evolution of
the coupling constant because the behavior of the entire
solution can be qualitatively understood by writing
($f_{mn}$ disappears from the formula in the case of
Wegner's equation)
\begin{equation}
\label{hmn}
H_{\lambda m n} \, = \, 
\sqrt{ E_m E_n } \, A_{mn} \, \exp{[-(E_m-E_n)^2/\lambda^2]} \, ,
\end{equation}
with matrix $A_{mn}$ of the form
\begin{equation}
\label{A}
A_{mn} = \delta_{mn} - g_\lambda - ih s_{mn} \, .
\end{equation}
The only term depending on $\lambda$ is the coupling 
constant $g_\lambda$; $h$ remains constant. 

Equation (\ref{A}) is not precise. For example, using
perturbation theory in the coupling constant $g_\Delta$, one
can find other terms in $A_{mn}$, such as $ - 2 g_\Delta^2
(E_m - E_n)^2/\lambda^2$. In fact, one also finds other
functions of the subscripts $m$ and $n$ and the width
parameter $\lambda$, and none of these structures is present
in Eq. (\ref{A}). Numerical calculation shows (see below)
that these structures combine to produce the behavior of
matrix elements of $H_\lambda$ that is qualitatively
captured by Eqs. (\ref{hmn}) and (\ref{A}). All matrix
elements can be described using variables $q_{mn} = (E_m -
E_n)/\lambda$. When $\lambda$ decreases, $q_{mn}$ for fixed
$m$ and $n$ increases and the corresponding matrix element
falls off. The dominant fall-off pattern is approximately
reproduced by the simple Gaussian function that multiplies
$A_{mn}$ in Eq. (\ref{hmn}). Although for matrix elements
that have $q_{mn}$ much larger than 1 the details of
$A_{mn}$ are not right and the Gaussian fall-off factor in
Eq. (\ref{hmn}) is not exact, the off-diagonal matrix
elements with large $q_{mn}$ (large changes of energy in
comparison to $\lambda$) become equivalent to zero anyway.
The simple Gaussian factor provides the same result.
The diagonal matrix elements $D_m$ with $m$ corresponding to
$E_m \gg \lambda$ are equal to the eigenvalues $\omega_m \gg
\lambda$ of the initial $H$. The RG evolution of the matrix
elements with $q_{mn} \sim 1$ was calculated numerically.
Eq. (\ref{A}) is accurate for matrix elements with $q_{mn}
\ll 1$. The smaller $q_{mn}$ the better the accuracy of Eqs.
(\ref{hmn}) and (\ref{A}). The key question here is how
$g_\lambda$ depends on $\lambda$. 

The best way to explain the behavior of $g_\lambda$ is to
use a generic example. On the basis of numerical experiments
with various choices for the parameter $b$, coupling
constants $g_\Delta$ and $h$, ultraviolet cutoff $\Delta$,
and infrared cutoff $E_M$, one representative case was
selected that clearly displays key features of the cycle of
the coupling constant $g_\lambda$ in effective theories. The
set of parameters includes $b = 4$, $g_\Delta = 0$, $h =
\tan{\pi/5}$, $N = 16$, and $M = -25$. The value of $b$ was
selected as a result of compromise between making it large
enough for a qualitative analysis (see next section) to be
sufficiently precise for numerical confirmation, and keeping
it small enough for the entire range of $\lambda$ to remain
within bounds set by requiring at least four digits of
numerical accuracy from 4th order Runge-Kutta integration
procedure for relevant matrix elements of $H_\lambda$.
$g_\Delta = 0$ as an initial condition was selected to keep
the calculation as simple as possible. Small values of
$g_\Delta$ will be discussed later in connection with
asymptotic freedom. The selected parameters are similar to
those used in Refs.
\cite{WilsonGlazekAustralia,GlazekWilsonAFBS} that analyzed
the case with $h=0$ (asymptotic freedom), to facilitate
comparison.

\begin{figure}[ht]
\includegraphics[angle=270,scale=0.34]{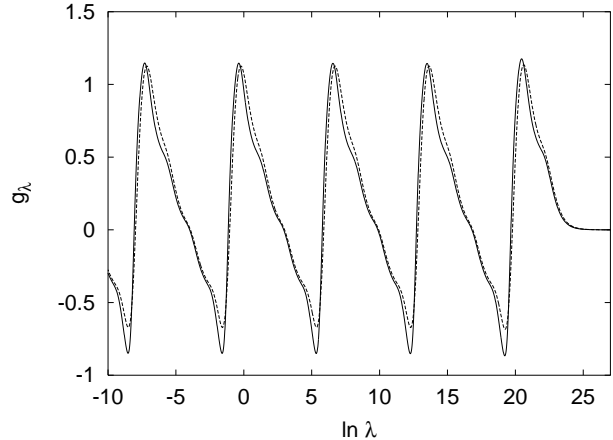}
\caption{The solid line displays the cycle. The 
broken line is an approximate result explained
in Section \ref{sec:change}.}
\label{fig:c5}
\end{figure}

Once the initial Hamiltonian is chosen, the similarity RG
transformation is fully determined by Eq. (\ref{srg}). The
running coupling constant is defined from the full solution
through
\begin{equation}
\label{gl}
g_\lambda = 1 - H_{\lambda M M}/E_M \, ,
\end{equation}
which is analogous to the definition of the fine structure
constant in the Thomson limit. The continuous line in Fig.
\ref{fig:c5} displays the resulting $g_\lambda$. The broken
line results from an approximate formula to be discussed in
the next section.

It is clear from Fig. \ref{fig:c5} that the effective
coupling constant oscillates with period $r = b^{\, p}$, where $p
= \pi/\arctan{(h)} = 5$. For example, one can read
the horizontal distances between successive points on Fig.
\ref{fig:c5} that correspond to $g_\lambda = 0.5$ and divide
them by $5 \ln{b}$. From the numbers that were used to draw
Fig. \ref{fig:c5}, one obtains (in the order from the
largest to lowest value of $\ln{\lambda}$ in the figure):
1.0326, 1.0032, 0.9996, 0.9999, 0.9995, 0.9998, 0.9994,
0.9998.

It is also clear that the pattern of change of effective
theories in a single cycle has two characteristic rates.
Namely, the coupling constant changes gradually over almost
the entire cycle and then suddenly jumps back to its initial
value. This is a generic feature. It was observed
numerically for all combinations of the parameters $1.1 < b
< 100$, $2 < p < 42$, and coupling constants $g_\Delta <1$,
with no indication of disappearing outside this range. For
example, if $p=15$, i.e., for $h = \tan{(\pi/15)}$, the
cycles are 3 times longer than in Fig. \ref{fig:c5}, but the
jump takes about the same distance along the horizontal
axis. For large $p$ that considerably exceeds $N-M+1$ and
corresponds to an extremely small skew symmetric imaginary part
in the interaction Hamiltonian, the cycle is so long that
one can hardly distinguish the cyclic behavior of the
coupling constant in the model with $N = 16$ from asymptotic
freedom (see Fig. \ref{fig:lcaf} in the next section). The
next section discusses the mechanism of the cyclic change
and shows how the cases of limit cycle and asymptotic
freedom are connected.

\section{ The cyclic change of $g_\lambda$}
\label{sec:change}

The pattern of incremental changes followed by a sudden big
jump that reverses the result of the gradual changes and
starts a new period of incremental variation, is a
characteristic feature of the model. The pattern is of
interest by itself and may provide useful analogies in
studies of other systems that display such behavior. 

Matrix elements of the Hamiltonian $H_\lambda$ can be written 
in the spectral form, 
\begin{equation}
\label{hpsi}
H_{\lambda m n} \, = \, \sum_{k = M}^N \omega_k \psi_{km}(\lambda) 
\psi^*_{kn}(\lambda) \; ,
\end{equation}
where $\omega_k$ is the $k$-th eigenvalue of the model and 
$\psi_{km}$ denotes $m$-th component of the corresponding 
effective wave function. The eigenstates are numbered from 
M to N in the order in which their eigenvalues appear on 
the diagonal when the similarity procedure is carried out 
all the way down to $\lambda=0$. The wave functions are 
normalized by the condition $\sum_{m=M}^N |\psi_{km}(\lambda)
|^2 =1$. The eigenvalues are independent of $\lambda$. 

The spectral decomposition implies through Eq. (\ref{gl}) 
that
\begin{equation}
\label{gpsi}
g_\lambda \, = \, 1 - {1 \over E_M} \,
\sum_{k = M}^N \, \omega_k \, \mathtt{a}_k(\lambda) \; ,
\end{equation}
where $\mathtt{a}_k(\lambda) = |\psi_{kM}(\lambda)|^2$ is
the probability of the lowest energy component in the
eigenvector number $k$ in the effective basis corresponding
to $\lambda$. Taking into account that in the flow of
$H_\lambda$ towards small $\lambda$ large eigenvalues appear
one after another on the diagonal and off-diagonal matrix
elements go to zero, one can conclude that the numbers
$\mathtt{a}_k(\lambda)$ are reduced to zero one after
another starting from large $k$. The eigenvector components
are expected to vary with $\lambda$ as the interaction does,
and the approximate Eq. (\ref{hmn}) suggests that
$\mathtt{a}_k(\lambda) \sim \mathtt{a} _k(\infty)
\exp{[-2(E_k - E_M)^2/\lambda^2]}$. A simple further 
consideration provides some insight concerning more 
precise description of the behavior of $\mathtt{a}_k(\lambda)$ 
as a function of $\lambda$. 

Namely, in the case of $N = M+1$, one can solve Eq.
(\ref{srg}) analytically and the solution says that the
small components of the eigenvectors indeed fall off as 
a Gaussian function, but the rate is given by $\exp{[-(\omega_k -
\omega_M)^2/\lambda^2]}$, i.e. the eigenvalues of $H_0$ are
replaced by the eigenvalues of $H$. Thus, knowing eigenvectors
and eigenvalues of $H$ one can write 
\begin{equation}
\label{al}
\mathtt{a}_k(\lambda)\, = \, c_k(\lambda) \, \mathtt{a}_k(\infty) 
\, \exp{[-2(\omega_k - \omega_M)^2/\lambda^2]} \, .
\end{equation}
$c_k(\lambda)$ is an associated normalization factor. It is
computed using the condition that the similarity 
transformation does not change the norm and assuming that
the RG evolution of the effective wave functions 
is given by the simple formula
\begin{equation}
\label{norm}
\psi_{km}(\lambda)\, = \, c_k(\lambda) \, \psi_{km}(\infty)
\, \exp{[-(\omega_k - \omega_m)^2/\lambda^2]} \, .
\end{equation}
Phases of the complex wave-function components $\psi_{km}(\infty)$
increase in steps with $m$ for $m > k$ with a period equal to the 
number of the steps needed to obtain $2\pi$. A similar pattern occurs 
in the three-body dynamics~\cite{MohrFurnstahlPerryWilson}.

\begin{figure}[ht]
\includegraphics[angle=270,scale=0.34]{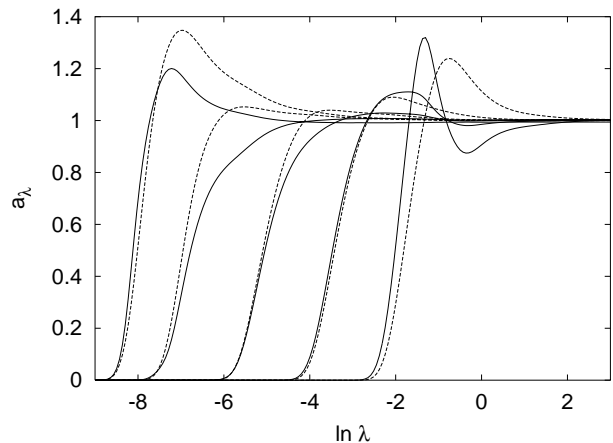}
\caption{ Full lines show the $\lambda$-dependence of lowest 
energy components, i.e., $\mathtt{a}_\lambda = \mathtt{a}_k
(\lambda)/\mathtt{a}_k(\infty)$, for 5 eigenvectors belonging 
to one cycle with $k = -6$, $-5$, $-4$, $-3$, and $-2$, from 
left to right, respectively, and the broken lines display 
results of the approximate Eq. (\ref{al}).}
\label{fig:wf}
\end{figure}

Eq. (\ref{hpsi}) with formula (\ref{norm}) for $\psi_{km}(\lambda)$ 
is now used to approximate the entire RG evolution of the 
effective Hamiltonian matrices obtained from solutions of 
Eq. (\ref{srg}): the effective wave functions evolve with
$\lambda$ through the exponential factors $\exp{[-(\omega_k 
- \omega_m)^2/\lambda^2]}$ and the normalization coefficients 
$c_k(\lambda)$. The fall-off pattern of the approximate Eq. 
(\ref{al}) is compared with exact numerically computed behavior 
of the wave functions in Fig. \ref{fig:wf} using the example of 
5 states forming a cycle.
  
A comment is required concerning the order of numbering of
eigenstates, which is related to why $b=4$ was chosen in the
example. The need for a comment stems from the intrinsic
complexity of the eigenvalue problem and RG equations. A
natural ordering of eigenvectors is found by tracing the
sequence in which eigenvalues appear on the diagonal when
$\lambda$ is lowered. It turns out that the ordering depends
on the size of $b$ when the other parameters in the model
are kept fixed. For example, there is a change in the
ordering of eigenvalues between $b = 2.15$ and $2.20$ in the
case considered here. The entire Hamiltonian dictates where
on the energy scale the negative bound-state eigenvalues are
to appear. In other words, there is one negative eigenvalue
per each and every quartet of successive positive
eigenvalues but the modulus of the negative eigenvalue can
wander across one of the positive eigenvalues on the
diagonal when $b$ is changed. $b=4$ was chosen to stay clear
of this feature near $b \sim 2.2$. But the interference
among the evolving wave functions is not avoided entirely
because the bound state wave functions spread coherently
over the basis states and have more than one significant
component. The interference is the origin of the difference
between the qualitatively correct Eq. (\ref{al}) and the
actual dependence of wave functions on $\lambda$, visible in
Fig. \ref{fig:wf}. In any case, all eigenvalues occur in
sequences of powers of $r=b^ {\, p}$. There are $p-1$ such
geometric sequences of positive eigenvalues and one such
geometric sequence of negative eigenvalues. The sequence of
negative eigenvalues is shifted on the logarithmic scale
with respect to the sequences of the positive eigenvalues.
Moduli of the negative eigenvalues are close to geometric
averages of the neighboring positive eigenvalues. It is
enough to find numerically where one of the moduli of
negative eigenvalues fits in among the positive eigenvalues
to locate moduli of all other negative eigenvalues on the
energy scale (a numerical pattern loses accuracy near
the cutoffs $M$ and $N$).

Using Eq. (\ref{gpsi}) with input from Eq. (\ref{al}), one
obtains the broken line shown in Fig. \ref{fig:c5}. There is
only a slight discrepancy between the broken line and the
actual cycle, which follows from the complex interference
pattern that is not fully captured. An additional benefit of
the approximate formula is that it allows one to study
behavior of the coupling constant for large $N$, $|M|$, $p$,
and $b$ approaching 1 with much less numerical effort than
the Wegner equation itself. Although it is easy to underflow 
or overflow computer accuracy in studies of the logarithmic
effects, some preliminary studies indicated that the effective
coupling constant $g_\lambda$ makes its rapid but smooth
transition at the scale of binding for large $p$ and $b$
approaching 1 without any significant increase in size. The 
maximal value of the coupling constant may even decrease when 
$b$ is lowered toward 1. The width of the transition region 
always appears to match the width of the bound-state wave 
function.

It is clear from Fig. \ref{fig:c5} that the basic features
of the cycle in behavior of $g_\lambda$ as a function of
$\lambda$ can be described in the following way. The
similarity RG flow eliminates interactions that change
energy by more than $\lambda$ and this forces eigenvectors
to have their significant components squeezed on the $H_0$
energy scale to the region of size $\lambda$ around the
corresponding eigenvalues. The eigenvalues appear on the
diagonal of $H_\lambda$ one after another and the eigenvectors 
end up contributing only to the diagonal matrix elements, being
eliminated from further evolution entirely. On the other
hand, the elimination of $p$ states of a full cycle brings
the low-energy matrix elements, and therefore also the
coupling constant, to the same values. A similar effect of
elimination of states was shown already in Refs.
\cite{GlazekWilsonExample,GlazekWilsonUniversality} using
the Gauss elimination procedure as the RG transformation. In
fact, elimination of $p$ high-energy states produces a
cyclic coupling constant behavior because
\begin{equation} 
\label{sum} 
\sum_{k = n-p+1}^n \, \omega_k
\, \mathtt{a}_k(\infty) \, = \, 0 \, . 
\end{equation} 
This identity is satisfied in the initial model Hamiltonian 
for $M \ll n \ll N$ with great accuracy independently of 
the RG procedure. The same pattern matters when similarity 
changes $\mathtt{a}_k (\infty)$ to $\mathtt{a}_k (\lambda)$ 
and low energy components of high-energy states 
are eliminated one after another by the Gaussian weight that 
is characteristic of Wegner's equation. Since there is only 
one bound state per cycle, one term in the sum of Eq. (\ref{sum}) 
is a negative of all the others in a cycle together. This is 
why the incremental change in the coupling constant induced 
by RG evolution of $p-1$ positive-eigenvalue eigenstates is 
undone by evolution of one negative-eigenvalue eigenstate. 

The actual width of the rapid transition of $g_\lambda$ as
a function of $\lambda$ must be related to the width of the
bound-state wave function on the energy scale (in quantum 
field theory, the scale of eigenvalues of $H_0$ corresponds 
to the scale of kinematic momentum variables). For large $b$, 
only a few components of the bound-state wave function have
significant size and the transition occurs over a few powers 
of $b$. When $b$ is lowered, more components matter. The width 
of the bound-state wave function must determine the size of 
the rapid transition region because all significant components 
of the bound-state must be eliminated in order to produce the 
bound-state eigenvalue on the diagonal. Since a simultaneous 
rescaling of the parameter $\lambda$ and all matrix elements 
of a cyclic Hamiltonian by $r$ recovers the same array of
matrix elements, the widths of successive bound states in 
the cycle also change like successive powers of $r$ when 
$\lambda$ is lowered. Thus, the width of the bound states 
that are eliminated when $\lambda$ passes the scale of
binding is constant on the logarithmic scale. The cyclic 
variation of $g_\lambda$ must therefore occur with a constant 
width of the rapid-fall-off region on the logarithmic scale.  

\begin{figure}[htb]
\includegraphics[scale=0.5]{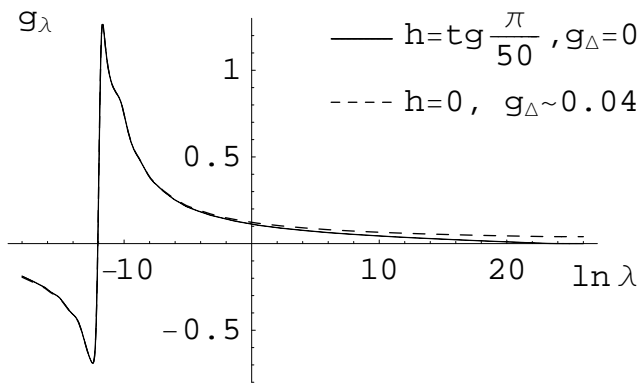}
\caption{ Behavior of $g_\lambda$ near the scale of binding 
in two cases: limit cycle and asymptotic freedom. The solid 
curve is obtained for the cycle with $p=50$ and the dashed curve 
for asymptotic freedom with $g_\Delta \sim 0.04$ (see text for
details).}
\label{fig:lcaf}
\end{figure}

When $p$ becomes very large, the incremental steps become
very small in size in comparison to the whole cycle (for
fixed $b$). But after the coupling $g_\lambda$ grows to
values much larger than $h$, its further change is governed
by the positive spectrum practically independently of $h$.
The same behavior of $g_\lambda$ for $\lambda$ near the 
scale of binding is obtained as in the case of asymptotic 
freedom in Refs. \cite{WilsonGlazekAustralia,GlazekWilsonAFBS}, 
where $h=0$. Fig. \ref{fig:lcaf} illustrates how $g_\lambda$ 
behaves in the two cases, both specified by $b=4$, $N=16$, $M=-25$.

The first case corresponds to a cycle with $h=\tan{(\pi/50)}$
and $g_\Delta=0$. In this case, there happens to be 
just one bound state with energy $E=-7.644479~10^{-6}$ (if
$N$ is increased to 66 and all other parameters are the
same, another bound state with eigenvalue $E'=-9.690529~
10^{24}$ is created in the first case and $E'=4^{50} E$). 
The second case corresponds to asymptotic freedom. The second
case also has just one bound state with the same binding 
energy but $h=0$ and $g_\Delta = 0.04000228$. In the second 
case, the bound-state is present due to a non-zero $g_\Delta$. 
It is visible in Fig. \ref{fig:lcaf} that as soon as $g_\lambda$
evolves toward decreasing values of $\lambda$ in the first case 
(limit cycle, solid curve) to values much larger than $h$, its 
further RG evolution toward still smaller values of $\lambda$ 
is governed by the size of $g_\lambda$ as in the second case 
(asymptotic freedom, dashed curve). 

The rapid transition in the dependence of $g_\lambda$ on
$\lambda$ when $\lambda$ passes through the scale of binding
corresponds to the well-known infrared singularity in QCD
(infrared slavery) that occurs when the momentum scale 
approaches $\Lambda_{QCD}$ in perturbation theory. In the
non-perturbative calculation in the model, the singularity
appears smoothed because it is spread over the width of a
bound-state wave function. 

The asymptotically free model is analogous to a cycle of a
very large range (very large $p$ for fixed $b$, which means
very small $h$) when $g_\Delta$ is set to a finite value in
the initial Hamiltonian so that the RG evolution brackets
the sudden-jump area on the energy scale due to a bound state.
If $g_\Delta \neq 0$ in the cyclic model, the spectrum is
slightly changed. It is shifted on the logarithmic scale and
the range of the cyclic variation of the coupling constant
is changed. The qualitative features for scales near the range 
of binding remain the same. 

The key difference between the cycle and asymptotic freedom
shows up when the coupling constant $g_\lambda$ approaches
0. In the case of asymptotic freedom, the coupling tends to
0 as a fixed point when $\lambda$ increases toward infinity.
In the case of a cycle, when $g_\lambda$ drops in size below 
$h$, the latter takes over in the dynamics and the
evolution toward larger $\lambda$ is altered.

\section{Discussion}
\label{sec:discussion}

Limit cycles of realistic effective quantum theories that
originate from models with divergences are difficult to
study because of their complexity. The complexity can also
obscure an underlying RG structure that may be close to a
limit cycle, as occurs in a number of cases mentioned in the
Introduction. The model discussed here is so simple that the
cycle of effective theories in it can be exposed and
analyzed without great effort. The study shows in a few
steps how the similarity RG procedure may be set up for
investigating limit cycles of effective theories. The cyclic
behavior of the coupling constant $g_\lambda$ is found by
solving Eq. (\ref{srg}) without calculating eigenvalues of
the model with large cutoffs. The structure of counterterms
is found using perturbative similarity RG transformation
order-by-order. The transformation guarantees that
eigenvalues of effective Hamiltonians with finite width
$\lambda$ are independent of $\lambda$. 

The same similarity RG procedure that works in the case of 
a limit cycle also works in the case of asymptotic freedom. 
The procedure shows how the two cases are related. It also 
produces an approximate structure of solutions of effective 
theories in the form of Eq. (\ref{norm}). This approximate 
structure works in both cases and thus appears to be more 
generally valid than the models used here to identify it. 
The meaning of this result can be described in the following 
way. 

Suppose a Hamiltonian for some physical system is proposed.
Suppose that the proposed Hamiltonian renders results that
are sensitive to arbitrary cutoffs that are not an intrinsic
part of the proposal but a technical element of necessity:
the proposed interactions spread states all over the space
considered in the theory up to the cutoffs and the model
without cutoffs, strictly speaking, does not exist. Matrix
elements of the model Hamiltonian can be re-interpreted
according to Eqs. (\ref{hmn}) and (\ref{A}) as corresponding
to an effective theory. The re-interpretation consists of
associating the basis states with energies known from
experiments and multiplying the model wave functions for any
given cutoff with a form factor analogous to the Gaussian
factor present in Eq. (\ref{norm}). The form factor contains
$\lambda$. For some finite value of $\lambda$ on the order
of observables of interest, the form factors in the new
(effective) Hamiltonian and in the wave functions eliminate
sensitivity to the arbitrary technical cutoffs. The
effective theory needs adjustment of constants at some
suitable value of $\lambda$. 

In fact, a similar sequence of steps characterizes construction 
of effective theories in atomic, nuclear, or particle physics 
in which one introduces form factors to smear singular 
interactions. (The similarity RG evolution of $g_\lambda$ 
in Hamiltonians of this article can be precisely related
to the evolution of coupling constants in Hamiltonians of 
quantum field theories with form factors, e.g., see
Ref. \cite{gQCD}.) All that the similarity RG procedure 
provides here is a new context of looking at the construction 
of effective theories, especially in the case of limit cycles.
In this context, when the energy {\it changes} described by the
effective Hamiltonians increase, new bound states of smaller
sizes (the size corresponds to $1/\lambda$) become a part
of the effective dynamics. Of course, a bound state of a
very small size cannot be resolved in processes limited to
small changes of energy. There, it can only play the role of 
a new particle.  

The close connection between two models: one with a limit cycle 
of a large scale period $r=\exp{(\pi /h)}$  and another one with 
asymptotic freedom, exposed here by the similarity RG procedure,
provides further insights concerning a very large range of scales. 
It suggests that Hamiltonians with extremely small imaginary matrix 
elements (extremely small coupling constants analogous to $h$ after 
$\ln b$ is factored out) can produce a new generation of binding 
effects where asymptotic freedom alone predicts no new physics. The 
limit cycle structure thus appears to suggest a scenario in which a 
dynamical correlation can emerge across a very large range of scales 
due to an extremely small coupling constant in a well-defined theory; 
the smaller the coupling constant $h$, the larger the scale period 
$r$. This is also a good reason for undertaking careful studies of 
limit cycles of effective theories using the similarity RG procedure 
and other means. 

{\bf Acknowledgment}

The author would like to thank Ken Wilson for numerous 
discussions concerning renormalization group and related 
issues.

\end{document}